# Title: Bulk Metallic Glasses Deform via Slip Avalanches

Authors: James Antonaglia[1,‡], Wendelin J. Wright[2,3,‡], Xiaojun Gu[2], Rachel R. Byer[4], Todd C. Hufnagel[5], Michael LeBlanc[1], Jonathan T. Uhl, and Karin A. Dahmen[1,*]

**Affiliations:**

[1] Department of Physics and Institute of Condensed Matter Theory, University of Illinois at Urbana Champaign, 1110 West Green Street, Urbana, IL 61801

[2] Department of Mechanical Engineering, One Dent Drive, Bucknell University, Lewisburg, PA 17837

[3] Department of Chemical Engineering, One Dent Drive, Bucknell University, Lewisburg, PA 17837

[4] Department of Physics & Astronomy, One Dent Drive, Bucknell University, Lewisburg, PA 17837

[5] Department of Materials Science and Engineering, Johns Hopkins University, 3400 North Charles Street, Baltimore, Maryland 21218

‡ *These authors contributed equally to this work.*

* Correspondence to: dahmen@illinois.edu

**Abstract:**
**Inelastic deformation of metallic glasses occurs via slip events with avalanche dynamics similar to those of earthquakes. For the first time in these materials, measurements have been obtained with sufficiently high temporal resolution to extract both the exponents and the scaling functions that describe the nature, statistics and dynamics of the slips according to a simple mean-field model. These slips originate from localized deformation in shear bands. The mean-field model describes the slip process as an avalanche of rearrangements of atoms in shear transformation zones (STZs). Small slips show the predicted power-law scaling and correspond to limited propagation of a shear front, while large slips are associated with uniform shear on unconstrained shear bands. The agreement between the model and data across multiple independent measures of slip statistics and dynamics provides compelling evidence for slip avalanches of STZs as the elementary mechanism of inhomogeneous deformation in metallic glasses.**

**One Sentence Summary:**
We show that bulk metallic glasses deform via slip avalanches of "weak spots", by demonstrating agreement of new high temporal resolution measurements of the slip-statistics and dynamics with the predictions of a simple mean field model for plastic deformation.

**Main Text:**
We show here that slowly sheared metallic glasses deform plastically via slip avalanches of weak spots. The weak spots are shear transformation zones (STZs), which are collective rearrangements of 10-100 atoms [1].

During high temperature deformation of metallic glasses (close to the glass transition), STZs operate independently and the material flows homogeneously, in agreement with STZ theory predictions over several orders of magnitude of stress and strain rate [1,2]. At lower temperatures metallic glasses deform inhomogenously via intermittent slips on narrow shear bands [3]. At low strain rates, these slip events are manifested as sudden stress drops, called serrated flow. Analytical [4,5] and computational investigations [6,7,8] suggest STZ operation, but experimental support has been challenging because slip events are both fast (with millisecond durations) and highly localized (with thicknesses <1 µm) [3]. Here we report experimental results on the stress drop dynamics and statistics, finding excellent agreement with analytic model predictions for the slip avalanche statistics of weak spots or STZs.

Many other materials—including crystals and densely packed granular solids—exhibit sudden slips during inelastic deformation. Although the mechanisms of deformation differ, the statistics and dynamics of the slip events are described by the same simple mean-field model of plastic deformation [9,10]. The model assumes that weak spots slip and then restick whenever the local shear stress exceeds a local slip threshold. Weak spots in crystals are dislocations, while in a metallic glass they are STZs. Through elastic interactions a slipping weak spot can trigger others to slip creating a slip avalanche. In crystals the slip can locally strengthen the material, while in metallic glasses it weakens it (through dilatation from STZ operation). For weakening materials (such as metallic glasses) the model predicts two types of slip avalanches: (1) *small slip avalanches* with a power law size distribution and self-similar dynamics, and (2) less frequent but almost regularly-recurring *large slip avalanches* with crack-like scaling behavior [9,10]. The model predicts both the statistics and the dynamics of the slip avalanches. Its agreement with experimental observations provides strong evidence that shear banding in metallic glasses arises from the collective slips of STZs.

To test the model we conducted uniaxial compression testing of metallic glass specimens in a high-stiffness, precisely-aligned loading train with a fast-response load cell and high-rate data acquisition (Figure 1a and Supplementary Information (SI)). During compression the specimen deforms elastically until a shear band or slip event initiates. This causes the displacement rate to temporarily exceed the displacement rate imposed on the specimen, resulting in a stress drop (Figure 1b). The size of the stress drop is proportional to the slip size. Subsequently the stress increases until initiation of another slip event. We measure the stress drops with unprecedented high time-resolution to resolve the dynamics of the slips. This enables us to extract a wide range of predicted scaling exponents and scaling functions that uniquely identify the underlying slip statistics and dynamics.

We briefly summarize some of the model predictions for the stress drop statistics [9,10] and the agreement with the experimental data. The complementary cumulative size distribution C(S), (i.e., the fraction of stress drops larger than size S) scales as $C(S) \sim S^{-1/2}$ for the (small) avalanches spanning the power law scaling regime of sizes $S_{min} < S < S_{max}$ (Figure 2a). $S_{min}$ is the experimental noise threshold (visible in Figure 2c). $S_{max}$ is related to the low-frequency roll-off $\omega_{min}$ of the power spectrum, $P(\omega)$, which is the absolute square of the Fourier transform of the stress drop rate time series (see Figure 2d). The model predicts $P(\omega) \sim 1/\omega^2$ for frequency $\omega >> \omega_{min}$, and $P(\omega) \to$ constant for $\omega << \omega_{min}$. Furthermore $\omega_{min} \sim 1/T_{max} \sim 1/(S_{max})^{1/2}$, with upper duration limit $T_{max}$ and upper size limit $S_{max}$ of the scaling regime in Figures 2a-c. Figures 2a-d show agreement of experiments with predictions for C(S), the average avalanche durations T at size S, the maximum stress drop rate at size S, and $P(\omega)$ (before and after Wiener filtering, see SI) [9,10,11,12]. In each case the power law exponents of the observed serration statistics agree with the model predictions. We also observe agreement with eight other statistical measures, including predicted scaling *functions*, as discussed in Figures 3 and 4 and the SI. Although some power law distributions of slip *sizes* in metallic glasses have been studied [13], no model has previously been shown to agree with the many different statistical measures of Figures 2-4 and the SI, that include not only multiple predicted scaling exponents but also predicted scaling functions.

We also test the predicted dynamics of individual slip events, which are sensitive to key assumptions of the model. Adding inertial effects or weakening [14,15,16,17] or a delayed damping term [18] to the model changes its predictions for avalanche dynamics [17,18,19] even though they may (or may not) affect the critical exponents of the slip statistics (Figure 2).

The model predicts the temporal profiles of the small avalanches, i.e. the stress drop rate as a function of time averaged over all slips of similar duration T. Figures 3a and 3b compare the experimental profiles, scaled by their duration T, to the model predictions. The measured profiles look symmetric, as shown more clearly in Figures 3c, 3d, and 4a, in remarkable agreement with the simple mean field model that neglects both inertia and delay effects. As shown in Figure 4b models with inertial/weakening effects predict asymmetric profiles that are tilted to the right [18,19], while models with delayed damping effects predict profiles tilted to the left (as observed for Barkhausen noise and large earthquakes) [18,20,21,22]. In the inset to Figure 4c we show average avalanche profiles for four different avalanche sizes S (rather than durations T). The avalanche profiles collapse onto the predicted scaling function when both axes are rescaled by the predicted factor $S^{-1/2}$.

The comparison of scaling *functions* in Figures 3 and 4 is a much more stringent test of the model than any traditionally used discrete set of power-law exponents. The extensive tests of slip statistics (Figure 2 and SI) and slip dynamics (Figures 3 and 4) thus confirm that inhomogeneous deformation of metallic glasses proceeds via slip avalanches of STZs. Furthermore, the data are consistent with the model prediction that inhomogeneous deformation of metallic glasses is an ordinary (tuned) critical phenomenon with a limited scaling regime, as opposed to a self-organized critical phenomenon [13] (which would require self-similarity to apply to all avalanches, small and large). The results indicate that inertial and delay effects are negligible for the slip statistics and dynamics of the small avalanches.

The two types of avalanches (small and large) predicted by the model and observed in the experiments correspond to different modes of shear propagation (Figure S3 in SI). Small slips result from progressive deformation, i.e. a propagating shear front with little or no slip occurring behind the front. Large slips are akin to a mode II crack with uniform sliding along the entire shear plane [9,10]. In the experiments, the small slips occur when a shear band nucleates at a stress concentration (such as the specimen/platen interface or an internal pore) and propagates as a front away from the concentration into a region of lower stress. The reduced driving stress causes the shear band to arrest after only a limited amount of slip and a correspondingly small stress drop. The large stress drops occur when a shear band manages to span the specimen, allowing continued simultaneous shear deformation on a plane at (approximately) 45° to the loading axis [23,24]. Slip stops when the shear stress drops below a critical stress at which point the shear band arrests. This agrees with the model prediction of crack-like scaling behavior of the large slip avalanches [9,10].

As assumed in the model, structural disorder and the presence of defects cause the critical stresses at which slips initiate or stop to vary throughout the specimen. For both large and small avalanches, dilatation during STZ operation leads to a decrease in viscosity of the material in the shear band. This weakening is the key tuning parameter of the model [9,10] that determines the size of the scaling regime for the small avalanches.

In summary, high temporal-resolution experiments on the slow compression of metallic glasses for the first time simultaneously measure slip statistics and slip dynamics. The results agree with predictions of a simple mean field model [9,10]. This agreement implies that inhomogeneous deformation in bulk metallic glasses proceeds via slip avalanches of weak spots, as assumed by the model, and provides compelling experimental evidence for the importance of shear transformation zones in the initiation and operation of shear bands. The prediction and observation of two types of avalanches (small ones marked by scaling behavior and large ones above the scaling regime) is associated with two modes of shear band operation. Small slip events correspond to nucleation and propagation of shear fronts. Some small slips grow sufficiently large to exceed a critical stress and transition into large slip events with crack-like sliding across a mature shear band. The high information content contained in the serration statistics and the dynamic avalanche profiles allow discrimination among competing models, suggesting that similar experiments will provide new insights into deformation mechanisms of other materials [25].

**Materials and Methods:**
Three-millimeter-diameter rods of $Zr_{45}Hf_{12}Nb_5Cu_{15.4}Ni_{12.6}Al_{10}$ were prepared using arc melting and suction casting and verified to be amorphous with x-ray diffraction. Rectangular parallelepiped specimens 6 mm long with a cross-sectional area of 2 mm by 1.5 mm with tight dimensional tolerances (reported previously in [26]) were machined from the cast ingots. Quasistatic uniaxial compression tests were performed at a constant displacement rate and a nominal strain rate of $10^{-4}$ s$^{-1}$ using a screw-driven Instron 5584 in a high-stiffness, precisely-aligned load train as shown in Figure 1a. The stress data were acquired using a 60 kN Kistler piezoelectric load cell with a 180 kHz low-pass filter. The data were recorded using a Hi-Techniques Synergy P system at a rate of 100 kHz with a 40 kHz low-pass filter. The displacement data were acquired using an Epsilon Tech 3442 extensometer. Metallic glasses fail catastrophically under uniaxial compression with fracture propagating at speeds on the order of 170 m/s or faster [23]. For specimens of this size, the fracture event occurs over an elapsed time that is less than 12.5 μs. High-speed imaging confirms that the electronics of the system are unable to accurately track this rapid fracture event due to the presence of the low-pass filters. The fracture event therefore functions as a unit impulse to the system, and the corresponding stress versus time data is used as the unit impulse response for the purposes of Wiener filtering (see the SI). The durations of the slip events ranges from 0.73 to 21.1 ms, while the applied stress is recorded every 10 μs; therefore, we record, on average, several hundred data points during each stress drop, allowing us to extract information about both the statistics of the stress drops as well as the dynamics of individual slip events.

**References**

1. Schuh, C. A., Hufnagel, T. C. & Ramamurty, U. Overview No.144 - Mechanical behavior of amorphous alloys. *Acta Mater.* **55**, 4067–4109 (2007).

2. Lu, J., Ravichandran, G. & Johnson, W. L. Deformation behavior of the $Zr_{41.2}Ti_{13.8}Cu_{12.5}Ni_{10}Be_{22.5}$ bulk metallic glass over a wide range of strain-rates and temperatures. *Acta Mater.* **51**, 3429–3443 (2003).
3. Greer, A. L., Cheng, Y. Q. & Ma, E. Shear bands in metallic glasses. *Materials Science & Engineering R-Reports* **74**, 71–132 (2013).
4. Argon A. S. Plastic deformation in metallic glasses. *Acta Metall. Mater.* **27**, 47–58 (1979).
5. Steif P. S., Spaepen F. & Hutchinson J.W. Strain localization in amorphous metals. *Acta Metall. Mater.* **30**, 447–455 (1982).
6. Homer, E. R. & Schuh, C. A. Mesoscale modeling of amorphous metals by shear transformation zone dynamics. *Acta Mater.* **57**, 2823–2833 (2008).
7. Falk, M. L. & Langer, J. S. Deformation and Failure of Amorphous, Solidlike Materials. *Annu. Rev. Condens. Matt. Phys.* **2**, 353–373 (2011).
8. Demkowicz M. J. & Argon A. S. Autocatalytic avalanches of unit inelastic shearing events are the mechanism of plastic deformation in amorphous silicon. *Phys. Rev. B* **72**, 245206 (2005).
9. Dahmen, K. A., Ben-Zion, Y. & Uhl, J. T. A simple analytic theory for the statistics of avalanches in sheared granular materials. *Nature Phys.* **7**, 554–557 (2011).
10. Dahmen, K. A., Ben-Zion, Y. & Uhl, J. T. Micromechanical Model for Deformation in Solids with Universal Predictions for Stress-Strain Curves and Slip Avalanches. *Phys. Rev. Lett.* **102**, 175501 (2009).
11. LeBlanc, M., Angheluta, L., Dahmen, K. A. & Goldenfeld, N. Distribution of Maximum Velocities in Avalanches Near the Depinning Transition. *Phys. Rev. Lett.* **109**, 105702 (2012).
12. LeBlanc, M., Angheluta, L., Dahmen, K. & Goldenfeld, N. Universal fluctuations and extreme statistics of avalanches near the depinning transition. *Phys. Rev. E* **87**, 022126 (2013).
13. Ren, J. L., Chen, C., Wang, G., Mattern, N. & Eckert, J. Dynamics of serrated flow in a bulk metallic glass. *AIP Advances* **1**, 032158 (2011).
14. Ramanathan, S. & Fisher, D. S. Onset of propagation of planar cracks in heterogeneous media. *Phys. Rev. B* **58**, 6026–6046 (1998).
15. Schwarz, J. M. & Fisher, D.S. Depinning with dynamic stress overshoots: Mean field theory. *Phys. Rev. Lett.* **87**, 096107 (2001).
16. Schwarz, J. M. & Fisher, D.S. Depinning with dynamic stress overshoots: A hybrid of critical and pseudohysteretic behavior. *Phys. Rev. E* **67**, 021603 (2003).
17. Fisher, D. S., Dahmen, K., Ramanathan, S. & Ben-Zion, Y. Statistics of Earthquakes in Simple Models of Heterogeneous Faults. *Phys. Rev. Lett.* **78**, 4885–4888 (1997).
18. Zapperi, S., Castellano, C., Colaiori, F., & Durin, G. Signature of effective mass in crackling-noise asymmetry. *Nature Phys.* **1**, 46–49 (2005).

19. Baldessari, A., Colaiori, F. & Castello, C. Average Shape of a Fluctuation: Universality in Excursions of Stochastic Processes. *Phys. Rev. Lett.* **90,** 060601 (2003).
20. Dahmen, K. A. Nonlinear dynamics: Universal clues in noisy skews. *Nature Phys.* **1**, 13–14 (2005).
21. Mehta, A. P., Mills, A. C., Dahmen, K. A. & Sethna, J. P. Universal pulse shape scaling function and exponents: Critical test for avalanche models applied to Barkhausen noise. *Phys. Rev. E* **65**, 046139 (2002).
22. Mehta, A. P., Dahmen, K. A. & Ben-Zion, Y. Universal mean moment rate profiles of earthquake ruptures. *Phys. Rev. E* **73**, 056104 (2006).
23. Wright, W. J., Byer, R. R. & Gu, X. J. High-speed imaging of a bulk metallic glass during uniaxial compression. *Appl. Phys. Lett.* **102**, 241920 (2013).
24. Song S. X., Wang, X. L. & Nieh, T. G. Capturing shear band propagation in a Zr-based metallic glass using a high-speed camera. *Scripta Mater.* **62**, 847–850 (2010).
25. Baro, J., Alvaro, C., Illa, X., Planes, A., Salje, E.K.H., Schranz, W., Soto-Parra, E. & Vives, E. Statistical Similarity between the Compression of a Porous Material and Earthquakes, *Phys. Rev. Lett*. **110**, 088702 (2013).
26. Wright, W. J., Samale, M. W., Hufnagel, T. C., LeBlanc, M. M. & Florando, J.N. Studies of shear band velocity using spatially and temporally resolved measurements of strain during quasistatic compression of a bulk metallic glass. *Acta Mater.* **57**, 4639–4648 (2009).
27. Papanikolaou, S., Bohn, F., Sommer, R. L., Durin, G., Zapperi, S. & Sethna, J. P. Universality beyond power laws and the average avalanche shape. *Nature Phys.* **7**, 316–320 (2011).
28. Salerno, K.M., Maloney, C.E. & Robbins, M.O. Avalanches in strained amorphous solids: does inertia destroy critical behavior? *Phys. Rev. Lett.* **109**, 105703 (2012).
29. Fisher, D. S. Collective transport in random media: from superconductors to earthquakes. *Phys. Reports* **301**, 113–150 (1998).

**Acknowledgments**:

We thank Braden Brinkman, Sean Clarke, Nir Friedman, and Matthew Wraith for helpful conversations. We thank NSF DMR-1042734 (WW), NSF DMR-1107838 (TH), NSF DMR-1005209, NSF DMS-1069224 (KD) and MGA for support.

**Figures and Captions:**

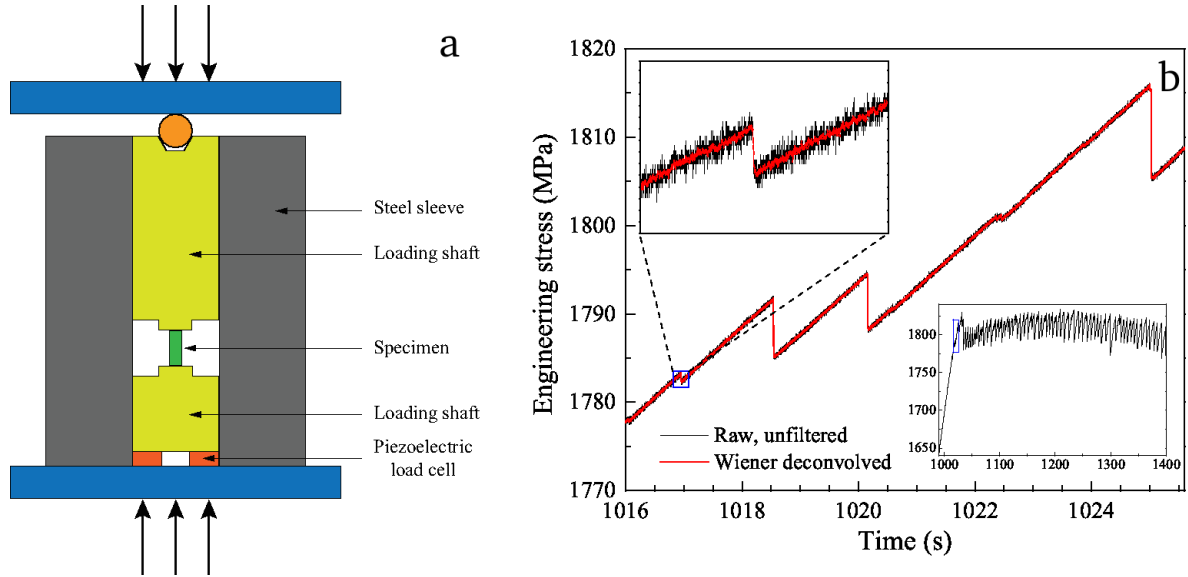

**Figure** 1 | Experimental setup and stress drops. **a,** Schematic diagram of experimental setup. Two tungsten carbide platens that are constrained by a steel sleeve compress the metallic glass specimen, see [26] for details. **b,** Lower-right inset: applied stress versus time. Main Figure: magnification of data in the small window in the lower-right inset. Slip-avalanches are manifest as sudden drops in applied stress. Upper-left inset is magnification from blue window in main figure, showing one stress drop. Black curves indicate raw, unfiltered stress time series, and red curves indicate the time series after Wiener deconvolution.

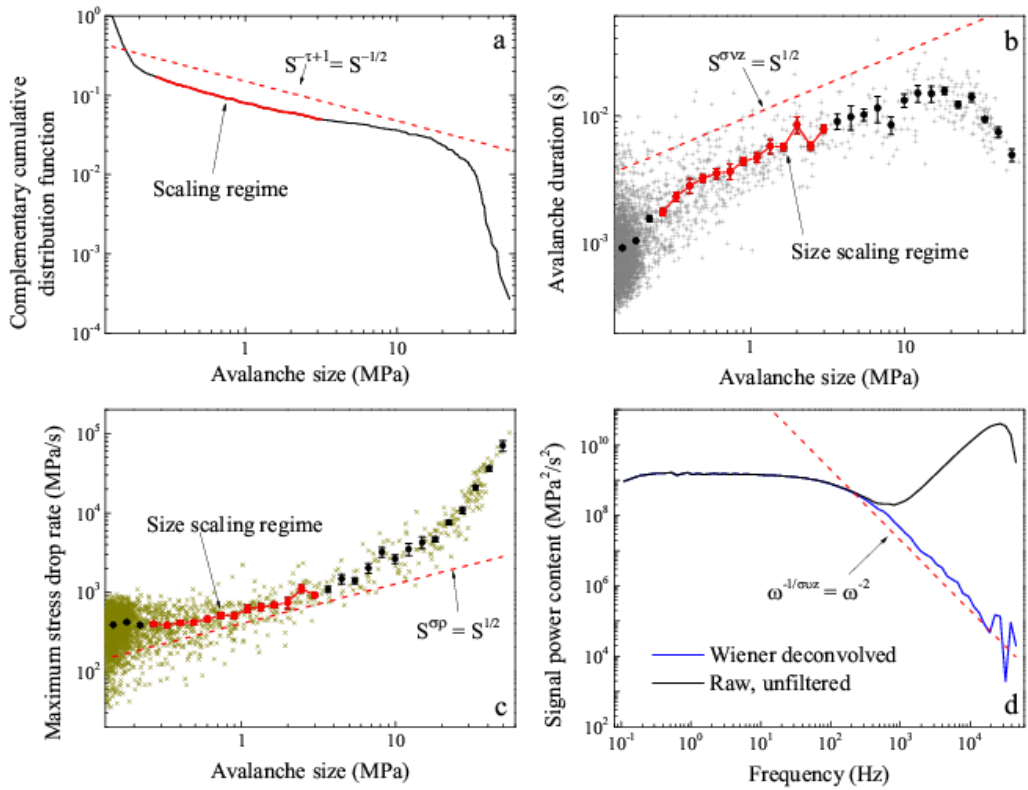

**Figure 2 |** Experimental Avalanche Statistics. Red dashed lines indicate model predictions [9,10,11,12]. **a,** Distribution of stress-drop sizes (3744 avalanches). The scaling-regime (red) extends from $2.6 \times 10^5$ Pa to $3.3 \times 10^6$ Pa, limited by high frequency noise (left) and large crack-like stress-drops (right). **b,** Avalanche duration $T$ versus size $S$. **c,** Maximum stress drop rates versus stress-drop size. **b and c**, One cross per avalanche. Black dots indicate average duration (b) or average maximum stress-drop rate (c) in one of 30 size bins. Error bars indicate 68% confidence interval. **d,** Power-spectrum of stress-drop rate time-series, before and after Wiener filtering.

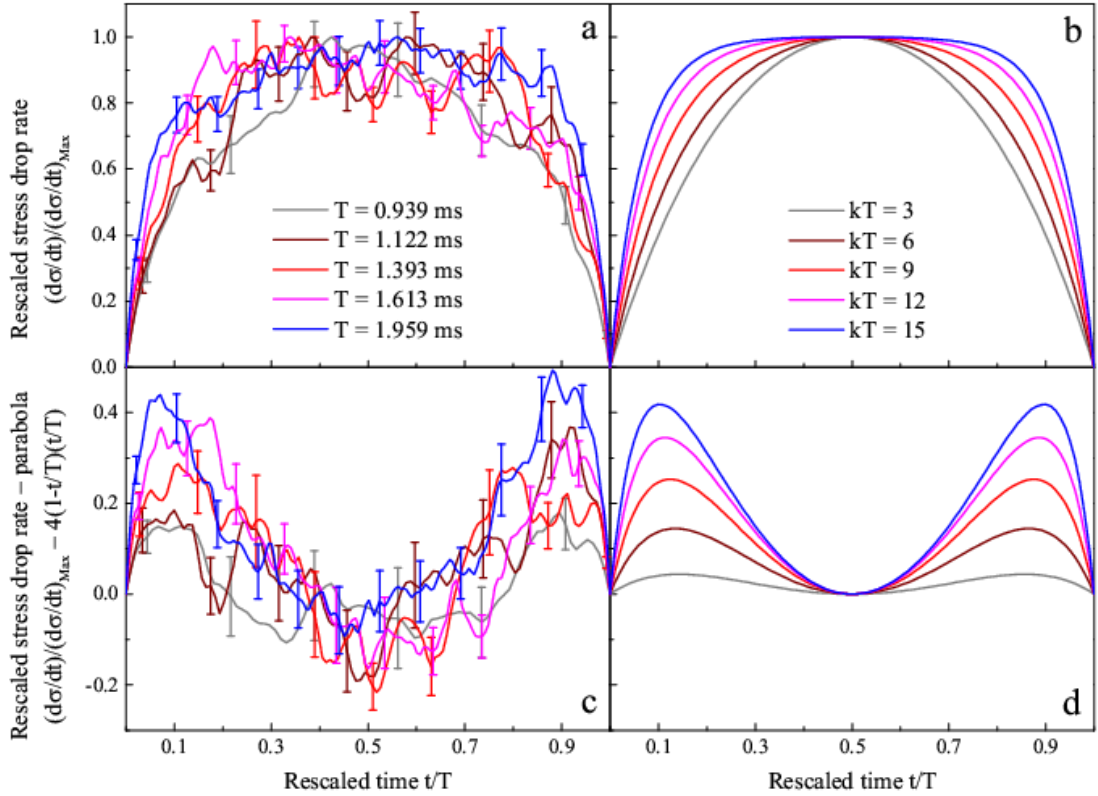

**Figure** 3 | Temporal avalanche profiles. **a,** Average stress-drop rate divided (normalized) by its maximum rate. Profiles are averaged over avalanches from small bins of their durations. Error bars reflect the variance of all avalanche profiles in each bin. **b,** Predicted average profiles from mean field theory. For increasing avalanche duration *T*, profiles become flatter, reflecting finite size effects, parametrized by *k* [27]. **c, d,** Difference between the normalized stress-drop rate and the parabolic form predicted for small avalanches. Larger avalanches deviate more from a parabola, consistent with the model prediction [27]. Note the clear symmetry of the profiles.

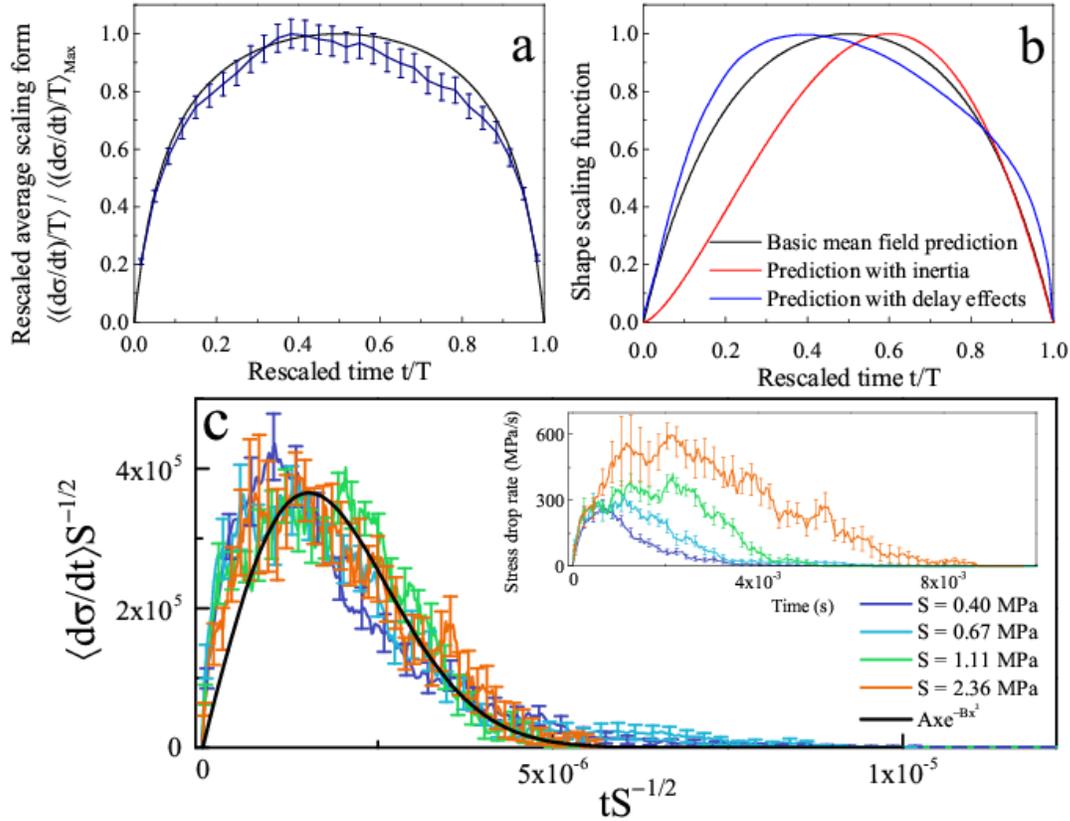

**Figure** 4 | Average avalanche profiles. **a,** Stress-drop rate profiles divided by duration T averaged over *all small* avalanches in the scaling regime. Agreement with the model prediction for fitting parameter k=1880±80 s$^{-1}$ (see SI). **b,** Model predictions for average avalanche profiles for different model assumptions [18,19,20,27]. **c,** Inset shows unscaled average stress-drop rate profiles for different stress-drop sizes S, collapsed in the main Figure (scaling both axes by a factor of S$^{-1/2}$). Agreement of collapse with predicted collapse scaling function (black line, for non-universal values of the constants A= 3.98×10$^{11}$ and B=2.18×10$^{11}$, see SI) [22,29].

# Supplementary Information:

**Calculation of avalanche properties**

Extraction of avalanche shapes from experimental data is straightforward when instrumental noise is small compared to the magnitude of avalanche sizes. The signal, in this case, applied stress, is differentiated over time, and when the derivative falls below zero (or a small negative threshold, in order to reduce the contribution of noise to avalanche distributions), the avalanche begins. When the derivative of the stress rises above zero or the small threshold, the avalanche ends; from this, the duration of the avalanche is computed. The magnitude of the avalanche is given by the difference in stress at these two times. The derivative of the stress between these two times constitutes the temporal shape of the avalanche.

The data for avalanches from two bulk metallic glass specimens of composition $Zr_{45}Hf_{12}Nb_5Cu_{15.4}Ni_{12.6}Al_{10}$ were combined to produce the analysis in this work. When analyzed individually, the results for the two specimens are the same (see Figure S1), but combining the data reduces the size of the error bars.

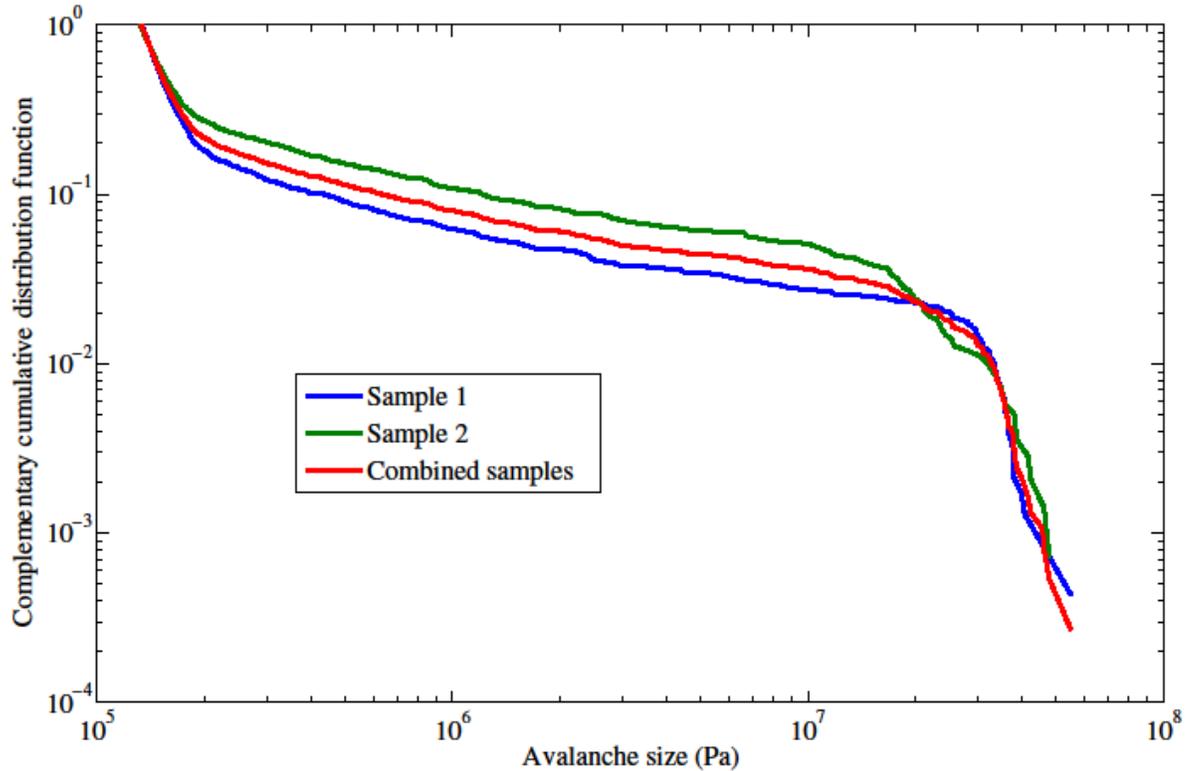

**Figure S1** | Complementary cumulative distribution of stress drop (avalanche) sizes for the two samples separately and for both samples combined. The figure shows that both samples produce essentially the same distributions with the same power law decay for the small avalanches (see also Figure 2a in the main paper).

### **Wiener filtering**

In order to identify when an avalanche begins and ends, it is important to properly filter out high frequency background noise, which may be of comparable amplitude to the stress drops caused by the slips. High frequency background noise can cause the fluctuations in the numerical derivative of the signal to be large. Wiener filtering has been successfully used to characterize avalanche profiles in previous experiments on Barkhausen noise [1].

Given a signal with added noise, which is then convolved with some impulse response of the measurement device, it can be shown that the Wiener filter minimizes the expected value of the discrepancy between the recorded signal and the true signal [2]. The filter is given in frequency space as

$$V_T(\omega) = \frac{V_E(\omega)}{h(\omega)} \frac{|s(\omega)|^2}{|s(\omega)|^2 + |n(\omega)|^2}$$

where $V_T(\omega)$ is the best estimate of the true signal, $V_E(\omega)$ is the experimentally observed signal, $h(\omega)$ is the impulse response of the measurement system, $s(\omega)$ is a theoretical prediction for the true signal, and $n(\omega)$ is the noise. The filter acts to dampen the Fourier components of the observed signal where the noise is more powerful than the expected signal. The filter passes those frequencies for which the noise power is small compared to the expected signal power.

The noise floor is established by holding the stress constant on a bulk metallic glass specimen at 1800 MPa, the stress around which all avalanches were found. The predicted signal is unknown, but from mean field theory, the power spectrum $P$ of the time derivative of the stress for slips in the scaling regime should follow a power law of $P=P_0\omega^{-2}$. The constant of proportionality is non-universal, so it is estimated from the observed signal. The constant is chosen such that $|s(f_C)| = |V_E(f_C)|$, where $f_C$ is a frequency that is on the order of the reciprocal of the smallest avalanche duration. If $f_C$ is much higher than this characteristic frequency, then too much high frequency noise passes through the filter. If $f_C$ is too low, then small-duration avalanches are smeared out by the filter, and the size of the scaling regime of the avalanches is diminished.

Avalanches can be identified by inspection in a plot of stress versus time, and the smallest in duration were generally around 3 ms; in this study, $f_C$=3000Hz<1/(3ms) was used. A significant difference in the distributions of avalanche durations was not found if $f_C$ was larger, around 5000Hz, but an $f_C$ lower than 2000Hz did obscure many small avalanches.

The impulse response function $h(\omega)$ is estimated from the response of the system to the fracture event because fracture provides the shortest-duration event available. For the specimens of the size and composition under consideration, the recorded force on the specimen drops from 1810 MPa to 0 in 50μs, which, with a sampling rate of 100 kHz, constitutes five data points. Inclusion of the impulse response did not have a significant effect on the derivation of the avalanche shapes.

This Wiener filtering process was applied to the derivative of the stress, and the significant dampening of instrumental noise allowed accurate extraction of avalanche shape profiles. A portion of the Wiener-filtered data is shown in Figure S2.

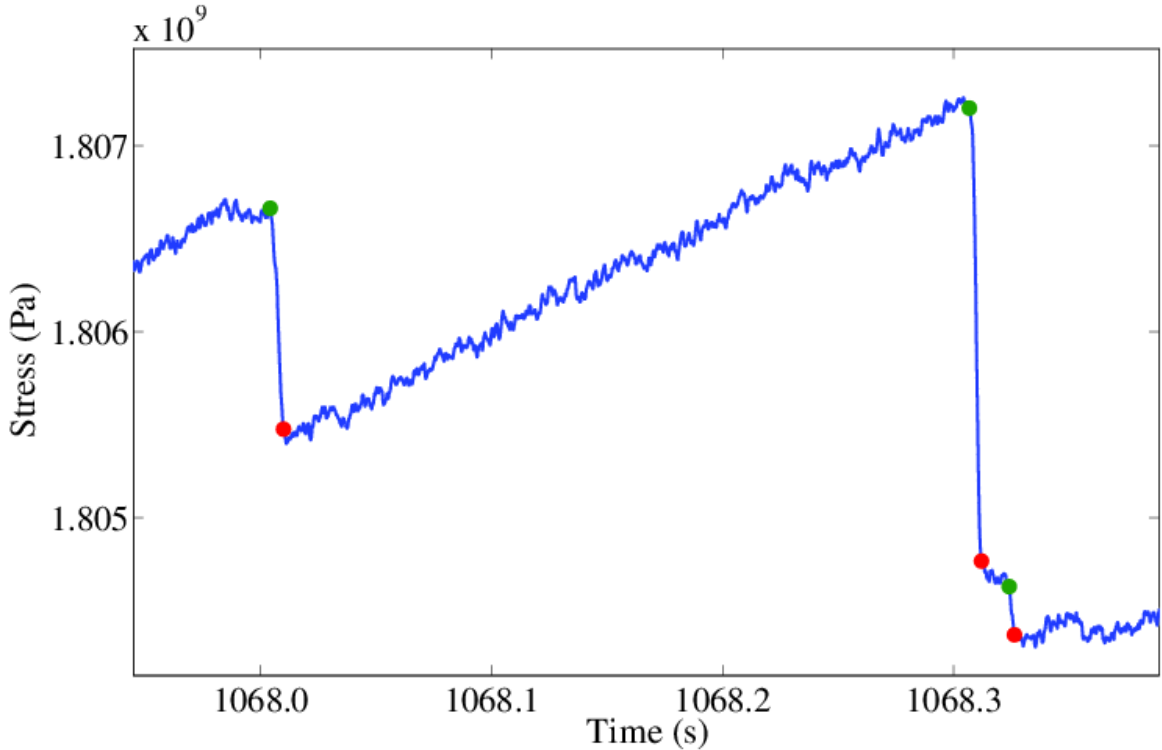

**Figure S2 |** After Wiener filtering, the derivative of the stress is reintegrated to give the filtered stress over time. The beginnings of avalanches are shown in green dots, and the ends are shown in red. The durations and sizes of the avalanches shown, in chronological order, are durations 5.7, 5.3, and 2.3 ms and sizes (stress drops) 1.2, 2.4, and 0.26 MPa, respectively.

**Criteria for locating scaling ("small") avalanches**

The size distribution (Figure 2a), the avalanche duration versus size (Figure 2b), the maximum stress drop rate of an avalanche versus its size (Figure 2c), and the power spectrum of the stress drop rate (Figure 2d) all show mean field power law scaling behavior [3,4,5,6,7,8,9,10] for the small avalanches. The breadth of the power law scaling regime is constrained by small-scale effects such as noise, large-scale effects such as weakening effects (predicted by the model [7,8,9,10]), and finite size effects [10]. In the following, we describe how the size of the scaling regime was established from the data. Figure 2b in the main text shows the duration of each avalanche plotted versus its size. Figure 2a from the main text shows that the complementary cumulative distribution C(S) of avalanche sizes S levels off for S > 3.3MPa from the predicted power law. This leveling off is predicted by the mean field model at the upper end of the scaling regime, so 3.3MPa is used as the upper limit of the scaling regime in the data. Figures 2b and 2c show that the lower limit is about 0.26MPa. Below this limit, strong noisy fluctuations are apparent. As shown in Figure S2, the Wiener filtering does not remove all noise. Avalanches below 0.26 MPa are on the same order as the remaining noise. (The noise fluctuations originate from medium-frequency Fourier components that were not removed by Wiener filtering.) Large avalanches, i.e., those that are significantly larger than the small avalanches in the scaling regime, have different scaling properties.

Both small and large slip avalanches can be explained intuitively in the context of the discrete mean field model of [7,8]. Weak sites in the material give way under stress and relieve that stress to other sites, triggering other weak spots to slip also. As the weak spots slip, they become weaker and can slip again during the same slip avalanche. The main difference between the small and large avalanches is that during the small avalanches, each weak spot slips on average at most once, as in a progressive slip. During the large avalanches, many weak spots slip many times, leading to simultaneous propagation of the slip avalanche (see Figure S3). Such simultaneous propagation significantly changes the scaling behavior of the large slip avalanches. The model predicts that for these large, simultaneous slips, the stress drop size S scales like a crack, as $S \sim A^{3/2}$, where A is the total slipping area, which is proportional to the number of different weak spots that slip in an avalanche [7,8,9]. In contrast, for the small slips in the scaling regime, the model predicts $S \sim A$ [7,8,9]. Similarly for the large slips, the largest stress drop rate $\langle (d\sigma/dt)_{max} | S \rangle$ for stress drops of size S is expected to scale as $\langle (d\sigma/dt)_{max} | S \rangle \sim S^{2/3}$ in the mean field model, while for the small slips in the scaling regime, we expect $\langle (d\sigma/dt)_{max} | S \rangle \sim S^{1/2}$. This behavior can be seen in the experimental data in Figure 2c in the main paper. Also, as expected, finite size effects skew the scaling for the very largest avalanches [7,8,9,10].

In agreement with the model predictions, the large crack-like events are less common than the small slips. Their extreme properties, such as their enormous stress drop rate (compared to the small slips) and their slightly asymmetric Gaussian-peak-like shape profiles (Figure S4), differ considerably from those of the small slips (Figures 3 and 4a) [7,8,9].

The scaling regime of size is given in Figure 2a, and the relationship between the sizes and durations is given in Figure 2b. Thus, the sizes of the scaling avalanches can be used to estimate the scaling regime of avalanche durations. The stress drops that span from 0.26 MPa to 3.3MPa have a minimum duration of 0.73 ms and a maximum duration of 21.1 ms.

Upon inspection of shape profiles of avalanches in the scaling regime that have large durations (e.g., Figure S5), we notice that the large avalanches have smaller stress drop rates than expected, which may reflect temporal merging of avalanches, and other effects, such as heating. Merging of avalanches would make it difficult to obtain accurate estimates of duration distributions and duration scaling regimes. If overlaps are likely, then the estimate for the largest avalanche duration $T_{max}$ in the scaling regime may be too high; in that case rather, the true scaling regime of the durations should be given by the bounds $f_{Min}$ and $f_{Max}$ of the scaling regime of the average power spectrum, which is not skewed by the temporal overlap of avalanches [11]. To reiterate, we expect that the true upper and lower bounds of the scaling regime for the durations are given by $T_{Max} \sim 1/f_{Min}$ and $T_{Min} \sim 1/f_{Max}$ of the scaling regime of the power spectra in Figure 2d. Conversely, if overlaps are negligible, the lower and upper limits of duration scaling can be used to estimate the upper and lower limits of the predicted frequency scaling region of the power spectra. Using the shortest and longest durations of avalanches that belong to the scaling regime of avalanche sizes in our experiment (see Figure 2b in the main paper), we find $1/T_{Min} = f_{Max} = 1852$ Hz and $1/T_{Max} = f_{Min} = 47$ Hz. For our experiments, the lower limit of 47 Hz is too low, reflecting that avalanches of maximum duration $T_{Max}$ likely consist of several overlapping avalanches. In contrast, the upper limit of 1852 Hz is consistent with the power spectrum in Figure 2d, in which the power law scaling of the power spectrum is evident up to about 2000 Hz. (We do not use the higher frequencies in our scaling analysis, because they may be affected by the Wiener filtering procedure that used a frequency cutoff of 3000 Hz). The agreement of the lower limit $T_{Min}$ of the duration scaling regime with the upper limit of the observed scaling regime of the power spectra is consistent with the fact that overlaps are expected to be negligible for the short avalanche durations [11].

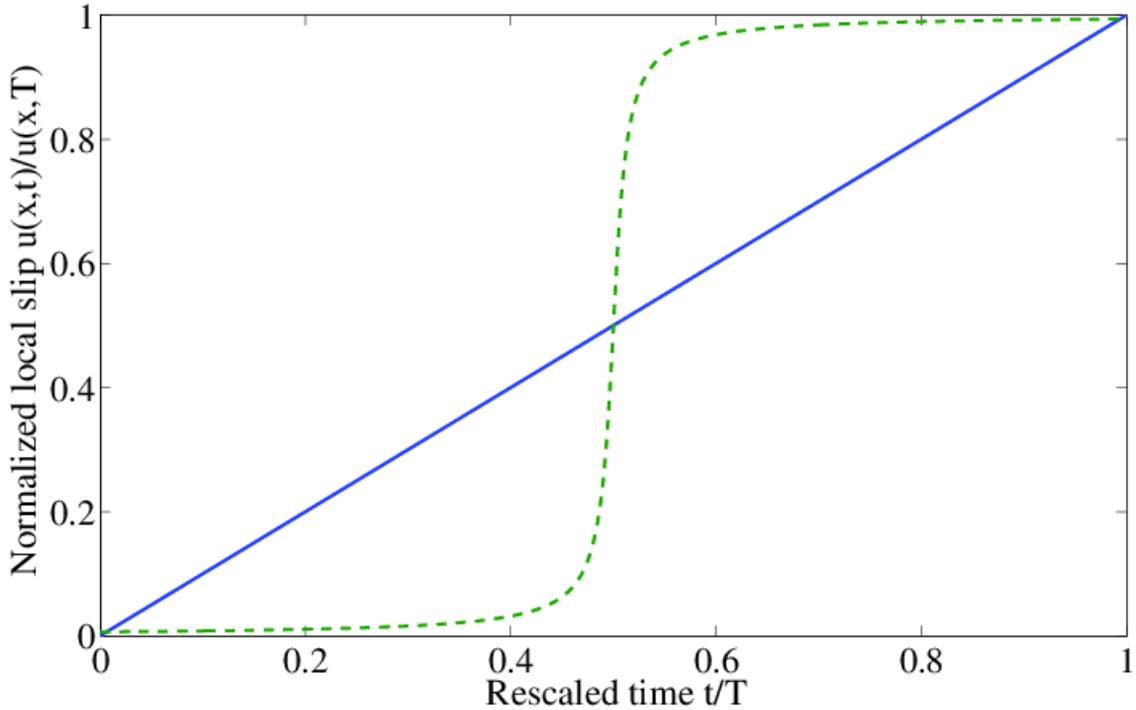

**Figure S3** | Simplified schematic of the local slip for small and large avalanches. The sketch shows the normalized local slip/displacement u(x,t) at a point x inside a slip avalanche as a function of time t during an avalanche that starts at time t=0 and finishes at time t=T. The green dashed line shows the time dependence of the normalized local slip for small slips, characterized by a propagating shear front with little or no slip occurring behind the front. The local slip

happens quickly, during a short time interval that is much shorter than the duration of the avalanche, as shown by the green dashed curve. In contrast, large slips show continual sliding along the entire shear plane, and once a local spot has started to slip, the local slip keeps increasing with time during the entire duration of the slip avalanche, as shown schematically by the blue line for the first spot to slip. Note that the details of the time dependence will vary from avalanche to avalanche; this plot is only a schematic to illustrate the key features of the different slips. Furthermore, the total slip $u(x,T)$ and the total duration T of the large avalanches represented by the blue curve is much larger than that of the small avalanches represented by the dashed green line. This fact is not visible in the above plot because both axes were rescaled by the respective total duration and total slip.

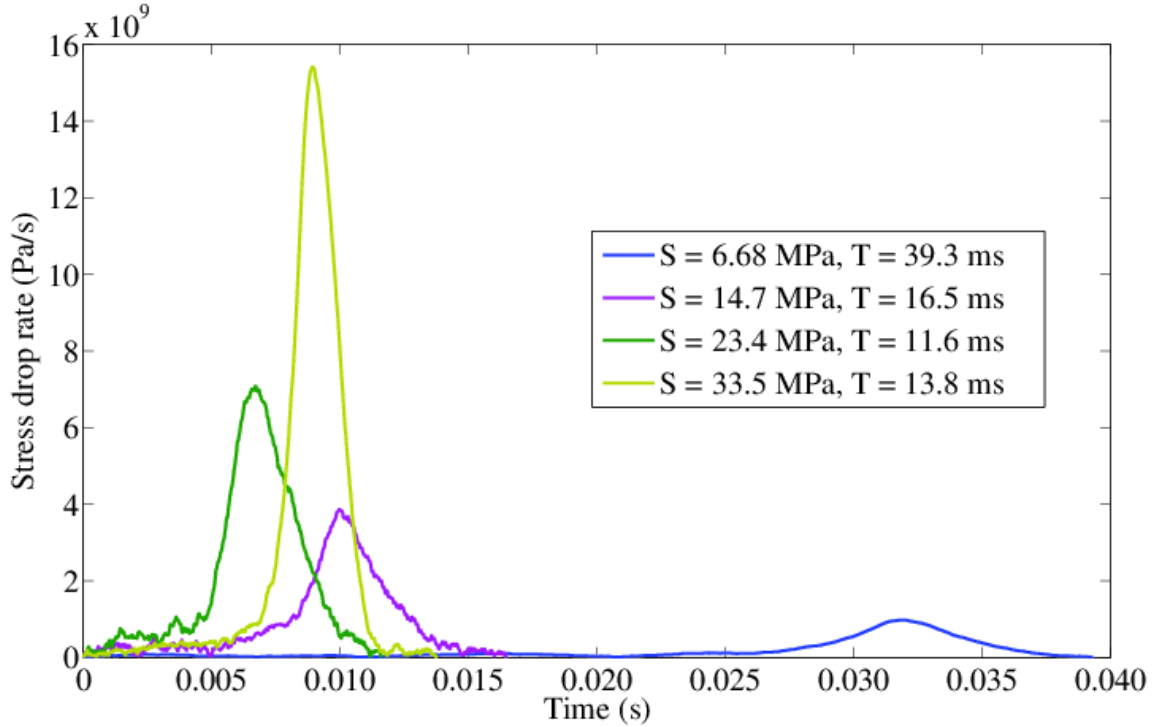

**Figure S4** | Examples of temporal avalanche profiles $<(d\sigma/dt)_{max}|S>$ of large avalanches. As predicted by the model, small avalanches in the scaling regime have stress drop rates that are much smaller than the large stress drop rates attained in these runaway avalanches. Only four are

shown here, but 183 avalanches were larger than 3.3 MPa, which is the upper region of size of the scaling regime.

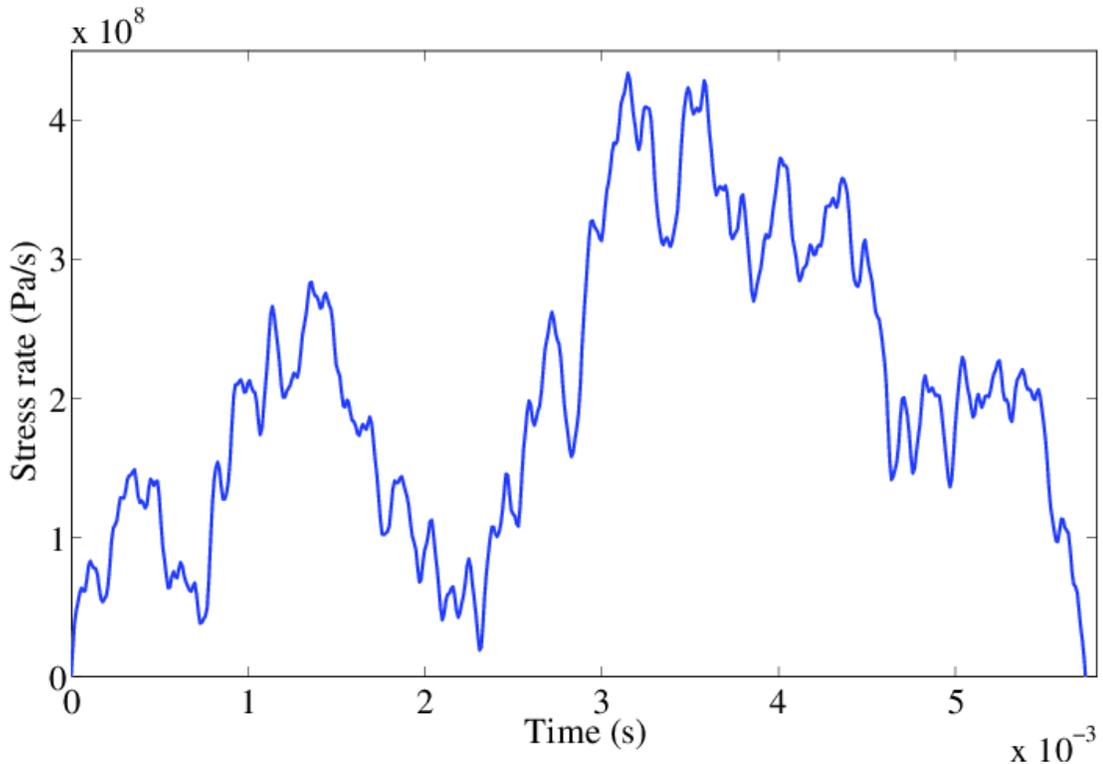

**Figure S5 |** Possible merging of avalanches. The size of this "small" avalanche is 1.19 MPa, so it falls into the upper half of the scaling regime. However, its duration T=5.73 ms is too long to be in the scaling regime of durations as suggested by the scaling regime of frequencies in Figure 2d, i.e. $T > 1/f_{Min}$, where $f_{Min}$ is the lower bound of the scaling regime of the power spectra in Figure 2d. The contradiction is explained by the fact that this avalanche may be composed of three avalanches that overlap in time. Consequently the combined event falls outside the true scaling regime of the durations that ranges from $T_{min}=1/f_{Max}$ to $T_{max}=1/f_{Min}$, where $f_{Min}$ and $f_{Max}$ are the boundaries of the scaling regime of the power spectra of Figure 2d.

**Method of averaging shapes**

To average many shapes of slightly varying durations as in Figure 3a, each avalanche is first scaled in time by its duration $T$ so that each avalanche extends from $t/T = \lambda = 0$ to $\lambda = 1$, where $t$ is time. Small bins of duration are taken, for example, centered around $T_1$ ranging from $0.85T_1 - 1.15T_1$. Then the time axis of each avalanche is averaged further into bins of $\lambda = t/T$, with each bin centered on one of the sample points of the shortest-duration avalanche in the duration bin. This ensures that each avalanche has the same number of sampling points, though they have different durations. From here, all avalanches with $0.85\,T_1 < T < 1.15\,T_1$ are averaged point by point at each point of $\lambda$.

Avalanches of similar size that are averaged without rescaling the time axis, i.e., those of the inset of Figure 4c, are simply summed, as the sampling rate of the experimental measurements is a constant 100 kHz. The summed shape is then divided by the number of avalanches sampled around a given size. Intuitively, this leads to a large bump close to $t = 0$ and then a long trailing tail caused by long-duration avalanches in the specimen [5,6].

The average of Figure 4a involves every avalanche in the scaling regime. Here, the interest lies in the universal shape profile, $V(\lambda, kT)$, which is plotted in Figure 3b for different values of the product of the duration $T$ with the parameter $k$, and again (after subtraction of the parabola that is predicted for $k = 0$) in Figure 3d. The model prediction for the stress drop rate $<\frac{d\sigma}{dt}|T>(t,T) \equiv \dot{F}(t,T)/a$, where $a$ is the cross-sectional area of the sample, and $\dot{F}(t,T)$ is the corresponding load drop rate, is given by [1].

$$<\frac{d\sigma}{dt}|T>(t,T) = TV(\lambda, kT)$$

$$V(\lambda, kT) = \frac{1}{kT}\frac{\left(e^{kT\lambda}-1\right)\left(e^{kT(1-\lambda)}-1\right)}{e^{kT}-1}$$

In Figure 4a, each avalanche is scaled in both time $t$ and stress drop rate $<\frac{d\sigma}{dt}|T>(t,T) = \dot{F}(t,T)/a$ by its duration $T$ and averaged over 30 evenly spaced bins of $\lambda = t/T$. The constant $k$ is tuned and used to fit to the resulting average of the experimental shapes.

**Experimental Observations**

      One of the two metallic glass specimens used for this analysis is shown in Figure S6. In this image, which was acquired during the compression test, the specimen is under stress and deforming plastically. One dominant shear band with a relatively large shear offset is clearly visible. Failure eventually occurred at this location. Scanning electron microscopy of the failed specimen reveals dozens of shear bands on the lateral surfaces; several are visible even at low magnification in Figure S7. It is likely that many shear bands were activated more than once (i.e., a single shear band produced one or more avalanches at different times [7,9,12]).

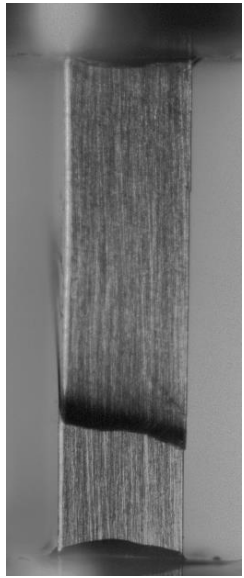

**Figure S6 |** A photograph of the metallic glass specimen used for this analysis. In this image, the specimen is under load. One dominant shear band is clearly visible. Some lubricant applied between the specimen and the loading platens can be seen at the specimen ends. The initial specimen dimensions in this view were 6.195 mm by 1.440 mm.

      Examination of the fracture surface in Figure S7 reveals the presence of a pore that is most likely due to a casting defect in the metallic glass ingot from which the specimen was machined. The pore is approximately 300 μm in diameter. The pore probably acted as a stress concentration that led to the repeated reactivation of the shear band that intersected it and ultimately to failure on this surface. The fracture surface does not intersect the ends of the specimen in contact with the loading platens. The vein pattern morphology that is characteristic of metallic glass fracture surfaces is apparent [13]. A pore was also observed on the fracture

surface of the second specimen. The ductilities of these two specimens (approximately 6% strain in both cases) may have been enhanced by the presence of the pores.

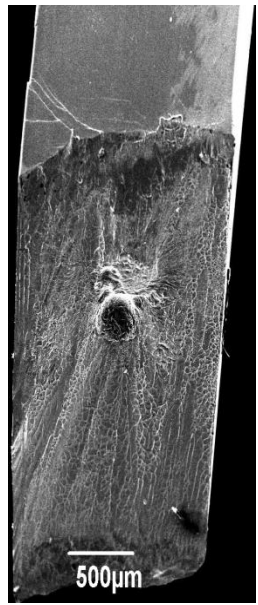

**Figure S7 |** A scanning electron micrograph of the fracture surface. Several shear bands are seen adjacent to the fracture surface. A 300 µm diameter pore due to a casting defect is visible.

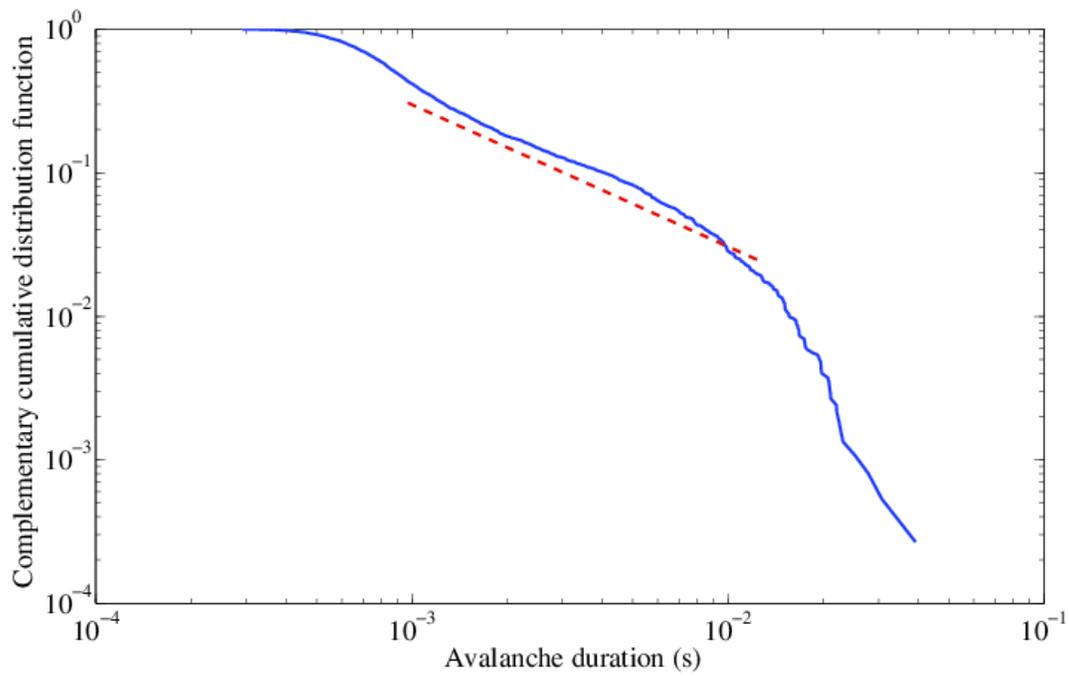

**Figure S8** | Complementary Cumulative distribution C(T) of stress drop durations T. The red dashed line indicates the power law $C(T) \sim T^{-1}$ which is predicted by the mean field model.

# List of the 12 observations on the avalanche statistics and their comparison to model predictions

The scaling regime in stress drop size S extends from $S_{Min} = 2.6 \times 10^5$ Pa to $S_{Max} = 3.3 \times 10^6$ Pa (see Figure 2a and main text). The power laws are fitted to the data for the small avalanches.

1. Complementary cumulative stress drop size distribution C(S). Model predicts $C(S) \sim S^{-(\tau-1)}$ for stress drop size S of the small avalanches (Figure 2a). Mean field theory predicts $\tau=1.5$, Power law fitted to experimental data $\tau-1=0.51 \pm 0.03$.

2. Duration T(S) versus stress drop size S, Model predicts with $\langle T(S) \rangle \sim S^{\sigma\upsilon z}$ shown in Figure 2b. Mean field theory predicts $\sigma\upsilon z=0.5$, Power law fitted to experimental data $\sigma\upsilon z=0.52\pm0.05$.

3. Maximum stress drop rate $\langle(-d\sigma/dt)_{max}|S\rangle$ during an avalanche of size S, see Figure 2c. Model predicts $\langle(d\sigma/dt)_{max}|S\rangle \sim S^{\sigma\rho}$ [10]. Mean field theory predicts $\sigma\rho=0.5$. Power law fitted to experimental data $\sigma\rho=0.47 \pm0.07$.

4. Power spectrum $P(\omega)$ of the stress drop rate $d\sigma/dt$, see Figure 2d. Model predicts $P(\omega) \sim \omega^{-1/\sigma\upsilon z}$.
   Mean Field theory predicts $1/\sigma\upsilon z = 2$. Power law fitted to experimental data $1/\sigma\upsilon z = 2 \pm 0.2$.

5. Temporal avalanche profiles at fixed duration T in Figure 3a have roughly the same shapes as the predicted mean field shapes in Figure 3b.

6. Temporal avalanche profiles at fixed duration are symmetric in time (Figure 3c), as predicted by the mean field model (Figure 3d), while other models predict asymmetric shapes (Figure 4b).

7. Average temporal profile at fixed T averaged over all small avalanches gives a shape that is very similar to the mean field shape (Figure 4a).

8. Temporal Avalanche profiles at fixed size S (Figure 4c) collapse onto the predicted *scaling function* shown by the black curve (with non-universal fitting parameters *A* and *B*), for the universal scaling exponents predicted by the mean field model.

9. The predicted rescaling of time *t* (*x*-axis) by $S^{\sigma\upsilon z}$ with the predicted mean field exponent $\sigma\upsilon z=0.5$ (the data collapse for $\sigma\upsilon z=0.5\pm0.05$).

10. The predicted rescaling of the slip rate (*y*-axis) by $S^{\sigma\rho}$ with the predicted $\sigma\rho=0.5$ (the

data collapse for $\sigma\rho=0.5\pm0.05$).

11. The complementary cumulative stress drop duration distribution $C(T)$ in Figure S8. Model predicts $C(T)\sim T^{-(\alpha-1)}$ with $\alpha=(\tau-1)/\sigma\upsilon z +1$ [11]. Mean field theory predicts $\alpha=2$, which is within the range power laws fitted to the data.

12. The model predicts small avalanches and large avalanches, with different shapes for the large avalanches than for the small ones. This agrees with the shapes obtained from the experimental data (see Figures 3, 4 and S4).

**References:**

[1] Papanikolaou, S., Bohn, F., Sommer, R. L., Durin, G., Zapperi, S. & Sethna, J. P. Universality beyond power laws and the average avalanche shape. *Nature* **7,** 316–320 (2011).

[2] Gershenfeld, N. *The Nature of Mathematical Modeling*. Cambridge 1999.

[3] LeBlanc, M., Angheluta, L., Dahmen, K. A. & Goldenfeld, N. Distribution of Maximum Velocities in Avalanches Near the Depinning Transition, *Phys. Rev. Lett.* **109**, 105702 (2012).

[4] Sethna, J. P., Dahmen, K. A. & Myers, C. R. Crackling Noise. *Nature* **410,** 242–248 (2001).

[5] Mehta, A., Dahmen, K. A. & Ben-Zion, Y. Universal mean-moment rate profiles of earthquake ruptures. *Phys. Rev. E* **73**, 056104/1-8 (2006).

[6] Fisher, D. S. Collective transport in random media: from superconductors to earthquakes. *Phys. Reports* **301**, 113–150 (1998).

[7] Dahmen, K. A., Ben-Zion, Y. & Uhl, J. T. Micromechanical Model for Deformation in Solids with Universal Predictions for Stress-Strain Curves and Slip Avalanches. *Phys. Rev. Lett.* **102**, 175501/1–4 (2009).

[8] Dahmen, K. A., Ben-Zion, Y., & Uhl, J. T. A simple analytic theory for the statistics of avalanches in sheared granular materials. *Nature Physics* **7**, 554–557 (2011).

[9] Fisher, D. S., Dahmen, K., Ramanathan, S. & Ben-Zion, Y. Statistics of Earthquakes in Simple Models of Heterogeneous Faults. *Phys. Rev. Lett.*, **78**, 4885–4888, (1997).

[10] LeBlanc, M., Angheluta, L., Dahmen, K. A., & Goldenfeld, N. Universal fluctuations and extreme statistics of avalanches near the depinning transition, *Phys. Rev*. E **87**, 022126/1–13 (2013).

[11] White, R. A. & Dahmen, K. A. Driving rate effects on crackling noise. *Phys. Rev. Lett*. **91**, 085702–1–4 (2003).

[12] Wright, W. J., Byer, R. R. & Gu, X. J. High-speed imaging of a bulk metallic glass during uniaxial compression. *Appl. Phys. Lett.* **102**, 241920 (2013).

[13] Leamy, H. J., Chen, H. S. & Wang, T. T. Plastic flow and fracture of metallic glass. *Metall. Trans*. **3**, 699 (1972).

**Author contributions**:
WW designed the experiments, and XG and RB contributed to the experimental design and performed the experiments. TH helped conceive the experiments and contributed to their interpretation. JA analyzed the experimental data and together with WW, TH, MLB, JTU and KD compared the results to the theoretical predictions. JA, WW, TH, JTU, and KD jointly wrote the manuscript. All authors read the manuscript and had opportunity to edit it.